\documentclass[twocolumn,showpacs,preprintnumbers,amsmath,amssymb,prb,dvipdfmx]{revtex4}

\usepackage{graphicx}
\usepackage{dcolumn}
\usepackage{bm}
\usepackage{color}
\usepackage{hyperref}
\usepackage{float}
\hypersetup{
colorlinks   = true,
urlcolor     = blue,
linkcolor    = blue,
citecolor   = blue
}

\begin{document}

\preprint{APS/123-QED}

\title{Magnetic moment of rare earth elements in $R_2$Fe$_{14}$B estimated with $\mu^+$SR} \unboldmath 

\author{Jun~Sugiyama$^{1}$}
 \email{juns@triumf.ca}
 \altaffiliation[Present address:]{\ CROSS Neutron Science and Technology Center, Tokai, Ibaraki 319-1106, Japan}
\author{Kazutoshi~Miwa$^1$}
\author{Hiroshi~Nozaki$^1$}
\author{Yuji~Kaneko$^1$}
\author{Bassam~Hitti$^{2}$}
\author{Donald~Arseneau$^{2}$}
\author{Gerald~D~Morris$^{2}$}
\author{Eduardo~J.~Ansaldo$^3$}
\author{Jess~H.~Brewer$^{2,4}$}

\affiliation{%
$^1$Toyota Central Research \& Development Laboratories~Inc.,
Nagakute, Aichi 480-1192, Japan
}%


\affiliation{
$^2$TRIUMF, 4004 Wesbrook Mall, Vancouver, BC, V6T 2A3 Canada 
}%

\affiliation{
$^3$Department of Physics \& Engineering Physics, University of Saskatchewan, 
Saskatoon, SK, S7N 5E2 Canada
}

\affiliation{
$^4$Department of Physics \& Astronomy, University of British Columbia,
Vancouver, BC, V6T 1Z1 Canada
}%

\date{\today}

\begin{abstract}
The ferromagnetic (FM) nature of Nd$_2$Fe$_{14}$B 
has been investigated with muon spin rotation and relaxation 
($\mu^+$SR) measurements on an aligned, sintered plate-shaped sample. 
A clear muon spin precession frequency ($f_{\rm FM}$) 
corresponding to the static internal FM field at the muon site 
showed an order parameter-like temperature dependence 
and disappeared above around 582~K ($\sim T_{\rm C}$). 
This indicated that the implanted muons 
are static in the Nd$_2$Fe$_{14}$B lattice 
even at temperatures above around 600~K. 
Using the predicted muon site and local spin densities 
predicted by DFT calculations, 
the ordered Nd moment ($M_{\rm Nd}$) 
was estimated to be {\color{black}3.31}~$\mu_{\rm B}$ at 5~K, 
when both ${\bm M}_{\rm Fe}$ and ${\bm M}_{\rm Nd}$ 
are parallel to the $\hat{\bf c}$-axis 
and $M_{\rm Fe}=2.1~\mu_{\rm B}$. 
Furthermore, $M_{R}$ in $R_2$Fe$_{14}$B 
with $R={\rm Y}$, Ce, Pr, Sm, Gd, Tb, Dy, Ho, Er, and Tm 
was estimated from $f_{\mu}$ values reported in earlier $\mu^+$SR work, 
using the FM structure proposed by neutron scattering 
and the same muon site and local spin density as in Nd$_2$Fe$_{14}$B.  
Such estimations yielded $M_{R}$ values consistent 
with those obtained by the other methods. 
\end{abstract}

\maketitle

\section{\label{sec:intro}Introduction}
Among many permanent magnet materials, Nd$_2$Fe$_{14}$B \cite{sagawa_84} 
and related intermetallic compounds \cite{herbst_91} 
are known to be very suitable for industrial applications, 
due to their high saturation magnetization ($M_{\rm s}=16~$kG), 
large energy product ($H_{\rm c}M_{\rm s}=64~$MGOe) 
and relatively low cost compared with that of Sm$_2$Fe$_{17}$N$_x$ \cite{coey_90}. 
Furthermore, 
although the Curie temperature ($T_{\rm C}$) is 592~K for Nd$_2$Fe$_{14}$B, 
the Nd$_2$Fe$_{14}$B phase does not decompose until 1428~K, 
resulting in flexibility of its synthesis process. 
Therefore, Nd$_2$Fe$_{14}$B and related compounds are widely used for 
high performance motors in many devices, electric vehicles and audio speakers. 
 
In the ferromagnetic (FM) phase, past neutron scattering measurements 
suggested a collinear spin structure at room temperature \cite{herbst_84}, 
in which both Fe and Nd moments ($M_{\rm Nd}$ \& $M_{\rm Fe}$) 
are aligned parallel along the [001] direction. 
The magnitude of the ordered $M_{\rm Fe}$ was almost saturated even at 300~K, 
{\color{black}i.e. $\sim2.2~\mu_{\rm B}$,} 
while $M_{\rm Nd}$ was initially thought to be below 1~$\mu_{\rm B}$ \cite{herbst_84}.  
{\color{black}The other neutron work reported that 
$M_{\rm Fe}\sim2.32(3)~\mu_{\rm B}$ and $M_{\rm Nd}\sim2.2~\mu_{\rm B}$ \cite{givord_85}, 
but the recent work revealed that 
$M_{\rm Fe}=1.9(1)~\mu_{\rm B}$ and $M_{\rm Nd}=1.5(1)~\mu_{\rm B}$ \cite{teplykh_13}.} 
Then, more detailed magnetization measurements at 4~K 
on $R_2$Fe$_{14}$B with $R={\rm La, Y, ...}$ 
revealed that $M_{\rm Fe}=2.1~\mu_{\rm B}$ \cite{herbst_91}, 
leading to $M_{\rm Nd}=3.2~\mu_{\rm B}$. 
In addition, Nd-NMR measurements suggested that 
$M_{\rm Nd}=2.7~\mu_{\rm B}$ at 4.2~K \cite{potenziani_85}. 
An X-ray magnetic circular dichroism (XMCD) study 
on $R_2$Fe$_{14}$B \cite{chaboy_96,soriano_00} 
implied that the ordered $M_{R}$s are very close to the values 
obtained from $gJ$ of $4f$ electrons, 
where $J$ is the quantum number of the total angular momentum 
and $g$ is the Land$\acute{\rm e}$ factor. 
This means that $M_{\rm Nd}\sim3.3~\mu_{\rm B}$. 
 
\begin{figure}[H] 
	\begin{centering} 
	\includegraphics[width=0.8\columnwidth]{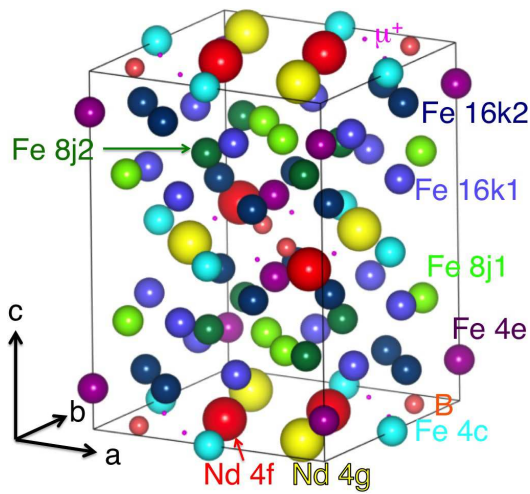} 
	\caption{ 
	The crystal structure of Nd$_2$Fe$_{14}$B in tetragonal symmetry with space group $P4_2/mnm$ 
	drawn by {\sf VESTA} \cite{vesta}.  
	Large red and yellow spheres show Nd at two different sites, 
	medium blue and green spheres show Fe at six different sites, and 
	small orange spheres show B. 
	Very small pink spheres represent the muon site (0.6744,0.8840,0) 
        predicted by first principles calculations (see text).  
	}  
	\label{fig:structure} 
	\end{centering} 
\end{figure} 
 
Furthermore, the FM spin structure in Nd$_2$Fe$_{14}$B 
was found to change at $135~{\rm K}(=T_{\rm SRT})$ 
due to a spin reorientation transition from 
a high-temperature phase with ${\bm M}\parallel[001]$ 
to a low-temperature phase with ${\bm M}$ canted along the [110] direction 
by magnetization measurements \cite{givord_84,abache_85,sagawa_85,hirosawa_86}. 
Initially, a collinear FM structure with a canting angle $\theta=30.6^{\circ}$ at 4.2~K 
was proposed based on magnetization measurements 
on a single crystal sample \cite{tokuhara_85}, 
where $\theta$ is the angle of ${\bm M}$ from the [001] direction to the [110] direction. 
However, both M$\ddot{\rm o}$ssbauer \cite{onodera_87} 
and XMCD \cite{chaboy_98} measurements suggested 
a non-collinear spin structure below $T_{\rm SRT}$. 
That is, $\theta_{\rm Fe}^{\rm M\ddot{\rm o}ss}=27^{\circ}$ 
and $\theta_{\rm Nd}^{\rm M\ddot{\rm o}ss}=58^{\circ}$ at 4.2~K, 
while $\theta_{\rm Fe}^{\rm XMCD}=28^{\circ}$ 
and $\theta_{\rm Nd}^{\rm XMCD}=40^{\circ}$ at 4.2~K. 
The continuation of XMCD work \cite{bartolome_00} indicated 
the formation of a further noncollinear spin structure 
among the Nd moments at temperatures between 80~K and $T_{\rm SRT}$, 
at which $\theta_{\rm Nd, 4f}\sim80^{\circ}$ and $\theta_{\rm Nd, 4g}\sim25^{\circ}$. 
 
In order to further elucidate the FM ground state of Nd$_2$Fe$_{14}$B, 
we need another technique sensitive to internal magnetic field(s) 
($\vec{H}_{\rm int}$) in solids. 
Unfortunately, neutron scattering is unlikely to be useful 
for investigating the magnetic nature of ferromagnets, 
because relatively weak magnetic diffraction peaks 
always overlap with strong nuclear Bragg peaks. 
{\color{black}Indeed, the estimated $M_{\rm Nd}$ with neutron ranges from 1 to 2.2~$\mu_{\rm B}$ 
\cite{herbst_84,givord_85,teplykh_13}, 
which is rather small compared with those obtained with the other techniques.} 
On the other hand, a positive muon spin rotation and relaxation ($\mu^+$SR) 
provides information on the local magnetic environments 
at the site(s) of the implanted muons, 
which usually locate at the interstitial site with the minimum electrostatic potential, 
regardless of magnetic order and/or disorder \cite{kalvius,yaouanc}. 
 
In fact, immediately after the discovery of the Nd$_2$Fe$_{14}$B system, 
a $\mu^+$SR experiment was performed 
at the Paul Sherrer Institut \cite{yaouanc_87,niedermayer_91} 
using powder $R_2$Fe$_{14}$B samples with 
$R=$ Y, Ce, Pr, Nd, Sm, Gd, Tb, Dy, Ho, Er, and Tm 
in the temperature range between 300 and 4.2~K. 
The $\mu^+$SR spectra obtained in zero external field (ZF) 
exhibited a clear oscillation with one precession frequency for all the samples, 
indicating both the formation of static FM order and a single muon lattice site. 
However, since it was very difficult to determine the correct muon site(s) in the lattice,  
the muon site was assumed to be a tetrahedral site 
with two Fe and two Nd nearest neighbors, 
based on the M$\ddot{\rm o}$ssbauer and neutron data 
of hydrated $R_2$Fe$_{14}$B \cite{ferreira_85,niedermayer_91}. 
In addition, the lack of information on the local spin density at the muon site 
made it eventually impossible to estimate the magnitude of $M_{R}$.  
As a result, the past $\mu^+$SR result is unlikely to be recognized 
as a crucial work for elucidating the magnetic ground state 
of Nd$_2$Fe$_{14}$B and $R_2$Fe$_{14}$B. 
 
We have therefore attempted to measure the $\mu^+$SR spectra for Nd$_2$Fe$_{14}$B 
up to above $T_{\rm C}$ to know the variation of $H_{\rm int}$ with temperature 
and to predict muon site(s) in the lattice with density functional theory (DFT) calculations. 
Using the predicted muon site and the measured local spin density at the muon site, 
the magnitude of $M_{\rm Nd}$ was clearly estimated even below $T_{\rm SRT}$. 
Furthermore, using the past $\mu^+$SR data for $R_2$Fe$_{14}$B and the predicted muon site, 
we have obtained a systematic change in $M_{R}$ with the number of $4f$ electrons in $R$.  
 
\section{\label{sec:exp}Experimental} 
 
Aligned sintered plates of Nd$_2$Fe$_{14}$B were prepared from 
jet-milled fine powder 
with the composition of 31.8Nd-0.98B-0.10Cu-0.90Co-0.15Al-0.05Ga-66.02Fe (wt\%).  
The mean particle size of the powder was about $6~\mu$m. 
The powders were then pressed under a magnetic field of 1.8~T 
followed by uniaxial pressing with 15~MPa. 
The pressed powders were sintered at 1293-1353~K 
for 4~hours in vacuum $(<10^{-2}~$Pa). 
Finally, the sintered powder $8\times8\times8~{\rm mm}^3$ cube 
was sliced into 1~mm thick plates 
with the aligned $c$-axis perpendicular to the plane. 
The preparation and characterization of the sintered sample 
are explained in more detail elsewhere \cite{takada_12}.  
 
\begin{figure}[tb] 
 \includegraphics[width=\columnwidth]{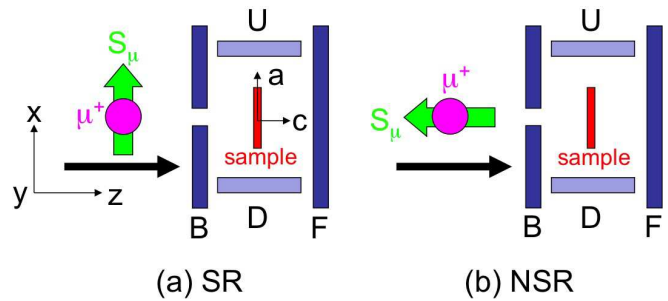} 
\caption{(Color online) 
Geometry of the $\mu^+$SR experiment in TRIUMF: 
four counters [backward (B), forward (F), up (U) and down (D)] 
detect decay positrons emitted 
in the $-z$, $+z$, $+x$ and $-x$ directions, respectively.  
The initial muon spin direction ${\bm S}_\mu(0)$ 
is in the $+x$ direction ($\parallel \hat{\bm a}$ of the plates) 
for spin-rotated (SR) mode (a) 
or in the $-z$ direction ($\parallel \hat{\bm c}$) 
for non-spin-rotated (NSR) mode (b).  
Thus if the internal magnetic field (${\bm H}_{\rm int}$) 
is parallel to $\hat{\bm c}$, 
only U and D counters will detect a muon spin oscillation, 
and that only in SR mode; 
but if ${\bm H}_{\rm int}\perp \hat{\bm c}$, 
only B and F counters in NSR mode will show an oscillatory signal.  
Using both configurations, 
one can estimate the magnetic anisotropy in the sample.  
} 
\label{fig:muSR} 
\end{figure} 
 
The $\mu^+$SR time spectra were measured 
on the {\sf M20} surface muon beam line 
using the {\sf LAMPF} spectrometer 
of the CMMS facility at TRIUMF in Canada. 
Four plates with $8\times8\times1~$mm$^3$ 
were arranged onto a sample holder 
with their $\hat{\bm c}$ axes parallel to the beam direction (${z}$) 
as defined in Fig.~\ref{fig:muSR}.  
For measurements in the $T$ range between 1.8 and 300~K, 
the samples were attached to a low-background sample holder 
in a liquid-He flow-type cryostat with 0.05~mm thick Al-coated Mylar tape. 
For measurements in the $T$ range between 300 and 600~K, 
the samples were fixed onto a silver plate by a $50~\mu$m thick titanium foil, 
which is sandwiched between a second silver plate 
with a $16\times16$~mm$^2$ square aperture through which incoming muons passed.  
For the former setup, there is essentially no background signal, 
while for the latter case the $\mu^+$SR signal 
naturally includes a background signal 
from muons stopped in the surrounding silver plate.  
 
The $\mu^+$SR spectra were obtained in either zero applied field (ZF) 
or transverse field (TF) 
with four positron detectors [backward (B), forward (F), up (U) and down (D)] 
arranged as shown in Fig.~\ref{fig:muSR}. 
The initial direction of the muon polarization [${\bm S}_\mu(0)$] 
relative to the plane of the plates 
was set by a Wien filter spin rotator. 
Here TF means the applied field is perpendicular to ${\bm S}_\mu(0)$, 
{\it i.e.} TF$\parallel y$ in this study. 
The experimental techniques are described in more detail elsewhere \cite{kalvius,yaouanc}. 
The resulting $\mu^+$SR 
data were analyzed with {\sf musrfit} \cite{suter_12}. 
 
The distributions of electrostatic potential and local spin density 
were predicted by DFT calculations 
with a generalized gradient approximation (GGA) 
plus on-site Coulomb interaction ($U$), 
as described in Sec.~\ref{sec:DFT}. 
 
\section{\label{sec:results}Results} 
\subsection{\label{sec:muSR}$\mu^+$SR} 
\begin{figure}[tbh] 
\begin{center} 
\includegraphics[width=70 mm]{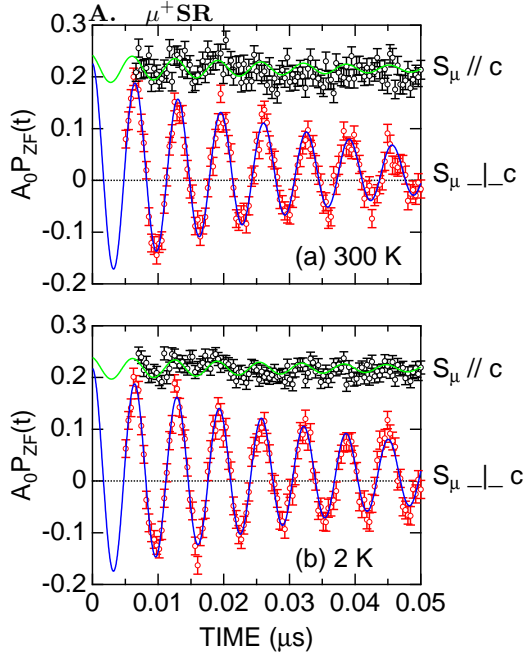} 
\caption{\label{fig:ZFspectra} (color online) 
The ZF-$\mu^+$SR spectrum for the sintered align Nd$_2$Fe$_{14}$B plate sample 
recorded at (a) 300~K and (b) 2~K  
in two different configurations:   
a non-spin-rotated (NSR) mode [${\bm S}_\mu(0)\parallel \hat{\bm c}$] 
shown in red 
and a spin-rotated (SR) mode [${\bm S}_\mu(0)\perp \hat{\bm c}$] 
shown in green. 
The solid lines represent the best fits using Eq.~(\ref{eq:ZFfit}). 
} 
\end{center} 
\end{figure} 
 
\begin{figure}[tbh] 
\begin{center} 
\includegraphics[width=0.8\columnwidth]{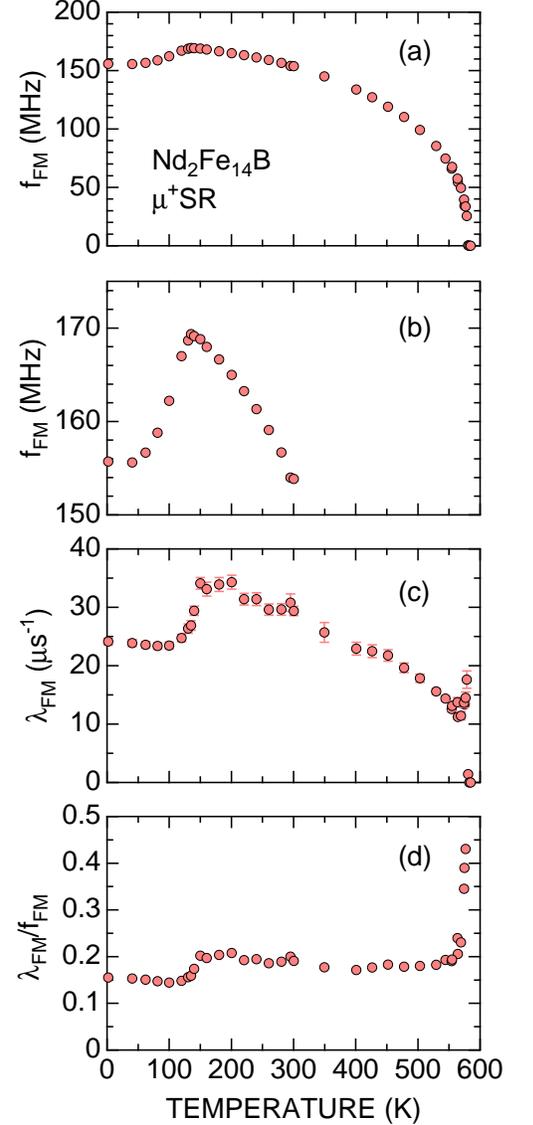} 
\caption{\label{fig:ZFana} (color online) 
The temperature dependences of 
(a) the muon spin precession frequency ($f_{\rm FM}$), 
(b) the magnification of the $f_{\rm FM}(T)$ curve to show the anomaly at around 135~K, 
(c) the exponential relaxation rate ($\lambda_{\rm FM}$), and 
(d) the ratio between $\lambda_{\rm FM}$ and $f_{\rm FM}$ 
for the Nd$_2$Fe$_{14}$B sample. 
The data were obtained by fitting the ZF-$\mu^+$SR spectrum with Eq.~(\ref{eq:ZFfit}). 
} 
\end{center} 
\end{figure} 
 
Figure~\ref{fig:ZFspectra} shows the ZF-$\mu^+$SR time spectra 
for the Nd$_2$Fe$_{14}$B sample recorded at 300 and 2~K. 
The spectrum obtained with SR mode [${\bm S_{\mu}}\perp\hat{\bm c}$] 
exhibits a clear oscillation, 
while that obtained with NSR mode [${\bm S_{\mu}}\parallel\hat{\bm c}$] 
shows mainly a non-oscillatory relaxation 
together with an oscillation with a very small amplitude. 
Since the Fourier transform frequency spectrum 
of the ZF-$\mu^+$SR time spectrum 
shows the presence of only one component, 
the two spectra were fitted by a combination of 
an exponentially relaxing cosine signal and 
an exponentially relaxing non-oscillatory signal: 
\begin{eqnarray} 
 A_0\, P_{\rm ZF}(t) &=& 
A_{\rm FM} \exp(-\lambda_{\rm FM}t) \cos(\omega_{\rm FM}t+\phi_{\rm FM}) 
  \cr 
  &+& A_{\rm tail} \exp(-\lambda_{\rm tail}t). 
\label{eq:ZFfit} 
\end{eqnarray} 
Here $A_0$ is the initial asymmetry, 
$P_{\rm ZF}(t)$ is the muon spin depolarization function in ZF, 
$A_{\rm FM}$ and $A_{\rm tail}$ are the asymmetries associated with the two signals, 
$\lambda_{\rm FM}$ and $\lambda_{\rm tail}$ 
are their exponential relaxation rates, 
$f_{\rm FM}(\equiv \omega_{\rm FM}/2\pi)$ is 
the muon Larmor frequency corresponding 
to the quasi-static internal FM field, 
and $\phi_{\rm FM}$ is the initial phase.  
At each temperature, the two spectra were fitted using 
common $\lambda_{\rm FM}$ and $f_{\rm FM}$. 
 
Such fits yielded 
$A_{\rm FM}^{S\perp c}=0.208(7)$, 
$A_{\rm FM}^{S\parallel c}=0.021(5)$, 
$\lambda_{\rm FM}=30.6(1.3)~\mu{\rm s}^{-1}$, 
$f_{\rm FM}=153.3(2)~$MHz,  
$\phi_{\rm FM}^{S\perp c}=1(2)~$deg, 
$\phi_{\rm FM}^{S\parallel c}=10(14)~$deg, 
$A_{\rm tail}^{S\perp c}=0.0159(2)$, 
$A_{\rm tail}^{S\parallel c}=0.2160(4)$, and 
$\lambda_{\rm tail}=0.0141(6)~\mu{\rm s}^{-1}$ at 300~K. 
Thus, the deviation from the $c$-axis 
of the magnetic {\color{black}field} at the muon site{\color{black}, i.e. the magnetic anisotropy at the muon site} is estimated to be 
$\Theta(300~{\rm K}) = 
 \tan^{-1}\left(A_{\rm FM}^{S\parallel c}/A_{\rm FM}^{S\perp c}\right)=7(4)~$deg.  
The same fit to the data at 2~K yielded $\Theta(2~{\rm K})=6(4)$~deg. 
This means that $\Theta$ is almost zero below 250~K within the accuracy of $\mu^+$SR. 
 
Figure~\ref{fig:ZFana} shows the temperature dependences of 
$f_{\rm FM}$, $\lambda_{\rm FM}$, and $\lambda_{\rm FM}/f_{\rm FM}$ 
for the Nd$_2$Fe$_{14}$B sample.  
The $f_{\rm FM}(T)$ curve exhibits an order parameter-like temperature dependence and 
$f_{\rm FM}$ disappears at temperatures above around 582~K ($=T_{\rm C}^{\rm \mu SR}$), 
which is slightly lower than $T_{\rm C}$ in literatures, 
{\it i.e.} 592~K \cite{sagawa_84,herbst_91}. 
Here it should be noted that $T_{\rm C}^{\rm \mu SR}$ 
is estimated from the data obtained in ZF, 
while the other techniques require the application of 
a large external magnetic field, 
which naturally enhances FM order. 
The $f_{\rm FM}(T)$ curve also shows 
a sharp local maximum at 135~K $(=T_{\rm SRT})$, 
indicating a change in the local FM environment 
caused by a spin reorientation transition.  
 
As temperature increases from 2~K, 
$\lambda_{\rm FM}$ decreases slightly up to 100~K, 
then suddenly increases up to 150~K, 
and then decreases again towards $T_{\rm C}$ 
with an increasing slope (d$\lambda_{\rm FM}$/d$T$).  
However, below the vicinity of $T_{\rm C}$, 
$\lambda_{\rm FM}$ rapidly increases with temperature, 
and then suddenly drops to zero at $T_{\rm C}$; that is, 
a critical behavior is observed below the vicinity of $T_{\rm C}$.  
 
It should be noted that $\lambda_{\rm FM}/f_{\rm FM}$, 
which corresponds to the normalized field distribution width, 
is almost temperature independent 
at temperatures below 100~K and 
at temperatures between 150 and 550~K. 
This means that 
besides the temperatures around a spin reorientation transition 
and below the vicinity of $T_{\rm C}$, 
$H_{\rm int}$ in the FM phase depends only on 
the magnitude of the ordered moments. 
These results suggest that muons are stable 
in the Nd$_2$Fe$_{14}$B lattice 
until $T_{\rm C}^{\rm \mu SR}$. 
The present result reproduces those in past $\mu^+$SR work 
carried out below room temperature 
\cite{yaouanc_87,niedermayer_91}. 
 
\begin{figure}[tbh] 
\begin{center} 
\includegraphics[width=0.8\columnwidth]{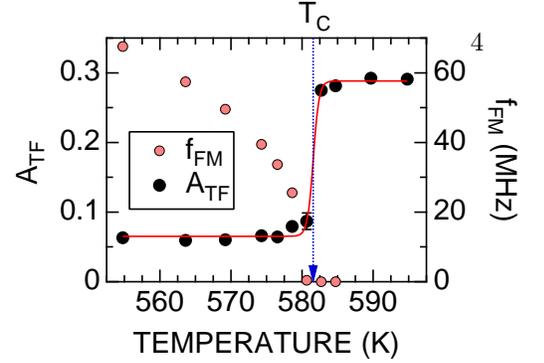} 
\caption{\label{fig:TFana} (color online) 
The temperature dependences of $f_{\rm FM}$ and 
the weak transverse field asymmetry ($A_{\rm TF}$) with TF=50~Oe.  
The $f_{\rm FM}(T)$ curve is the same as that in Fig.~\ref{fig:ZFana}(a). 
Here $T_{\rm C}$ corresponds to the midpoint 
of a step-like change in the $A_{\rm TF}(T)$ curve. 
} 
\end{center} 
\end{figure} 
 
In order to estimate $T_{\rm C}$ more correctly, 
Fig.~\ref{fig:TFana} shows the temperature dependence of 
the weak transverse asymmetry ($A_{\rm TF}$) measured with TF=50~Oe 
in the vicinity of $T_{\rm C}$, together with that of $f_{\rm FM}$. 
Here, ``weak" means that the applied TF is very small 
compared with $H_{\rm int}$ caused by FM order. 
The wTF-$\mu^+$SR spectrum was fitted by a combination of 
an exponentially relaxing cosine oscillation 
due to muon spin precession in TF and Eq.~(\ref{eq:ZFfit}): 
\begin{eqnarray} 
A_0\, P_{\rm TF}(t) &=& 
A_{\rm TF}\exp(-\lambda_{\rm TF}t)\cos(\omega_{\rm TF}t+\phi_{\rm TF}) 
\cr 
&+& A_{\rm FM} \exp(-\lambda_{\rm FM}t) \cos(\omega_{\rm FM}t+\phi_{\rm FM}) 
\cr 
&+& A_{\rm tail} \exp(-\lambda_{\rm tail}t). 
\label{eq:TFfit} 
\end{eqnarray} 
At temperatures $T\gg T_{\rm C}$, $A_{\rm FM}=A_{\rm tail}=0$; 
at temperatures $T\ll T_{\rm C}$, $A_{\rm TF}=0$. 
From the middle point of a step-like change in the $A_{\rm TF}(T)$ curve, 
$T_{\rm C}$ is estimated as 581.57(14)~K, 
because $A_{\rm TF}$ is proportional to 
the volume fraction of paramagnetic phases in a sample. 
The finite value of $A_{\rm TF}$ below $T_{\rm C}$ ($\sim0.06$) 
is from muons stopped in the surrounding silver plate.

\subsection{\label{sec:DFT}DFT calculations} 
 
First-principles calculations based on a 
density functional theory~(DFT)~\cite{Kohn1964a,Kohn1965a} 
have been performed to determine the muon site in ${\rm Nd_2Fe_{14}B}$. 
A self consistent field (SCF) calculation is carried out using the ultrasoft 
pseudopotential method\cite{Vanderbilt1990a,Miwa2011a}, 
where the on-site Coulomb interaction for localized Nd-$4f$ electrons 
is taken into consideration using 
the ${\rm DFT}+U$ method~\cite{Anisimov1995a,Miwa2018a}. 
The obtained pseudo SCF charge density is transformed 
into an all electron form 
with the projector augmented wave operators\cite{Blochl1994a}, 
from which the muon occupation site is estimated 
by the electrostatic potential analysis. 
{\color{black}
The program used for the DFT calculations is an original code developed by one of the authors (K.~M.), 
which has been successfully applied for various materials \cite{miwa_02,miwa_04,miwa_11,jun_13,miwa_18}.
}

The cutoff energies of plane waves are set to be 25 and 200~hatrees for 
the pseudo wavefunctions and the charge density, respectively. 
The $4\times 4\times 4$ $k$-point mesh is adopted for the Brillouin zone integration. 
The generalized gradient approximation~\cite{Perdew1996a} is used 
for the exchange-correlation functional. 
The effective Coulomb and exchange parameters for Nd-$4f$ orbitals are assumed to be 
$U=5$~eV\cite{alam_13} and $J=0.5$~eV, respectively. 
 
\begin{figure} 
\begin{center} 
\includegraphics[width=0.7\columnwidth]{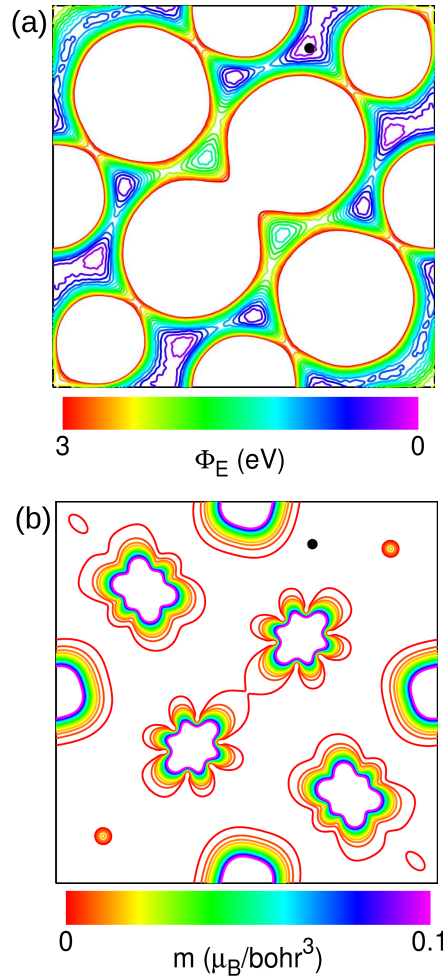} 
\end{center} 
\caption{ 
Contour plots for ${\rm Nd_2Fe_{14}B}$ in the (001) plane. 
(a)~Electrostatic potential $\Phi_\textrm{E}$ 
and (b)~spin density $m (= \rho^\uparrow - \rho^\downarrow$). 
The muon site is indicated by black circles. 
\label{fig:elpot}} 
\end{figure}

Table~\ref{table:str} shows the result of the structural relaxation in which 
atomic positions as well as lattice constants are fully optimized. 
The calculated parameters are in good agreement with the experimental ones~\cite{li_96}. 
Figure~\ref{fig:elpot}(a) depicts the electrostatic potential: 
The muon site is found to be $8i~(0.6745, 0.8838, 0)$ which is located near the center of 
a square base of a pyramid composed of Nd-3Fe-B atoms. 
As shown in Fig.~\ref{fig:elpot}(b), the spin density at the muon site is negligibly small, 
$\rho_{\rm spin} = -2\times 10^{-3}~\mu_B/\textrm{bohr}^3$, 
which is eventually zero. 
{\color{black}
It should be noted that the DFT calculations with $U=0$ provides very similar muon site and local spin density 
to those predicted with $U=5~$eV. 
This means that the two significant parameters, 
i.e. the muon site and  $\rho_{\rm spin}$, are not sensitive to $U$ 
in the Nd$_2$Fe$_{14}$B lattice. 
}
 
\begin{table}[htp] 
\caption{\label{table:str} 
Crystallographic parameters of ferromagnetic ${\rm Nd_2Fe_{14}B}$. 
Space group: $P4_2/mnm$~(No.~136). 
Lattice constants: $a=8.797$~{\AA}, $c=12.149$~{\AA}~(Calc.), and 
$a=8.795$~{\AA}, $c=12.188$~{\AA}~(Expt.). 
} 
\begin{ruledtabular} 
\begin{tabular}{ccccccccc} 
 & & \multicolumn{3}{c}{Calc.} & & \multicolumn{3}{c}{Expt.\footnotemark[1]} \\ 
 \cline{3-5} \cline{7-9} 
 & site & $x$ & $y$ & $z$ & \hspace*{1ex} & $x$ & $y$ & $z$ \\ 
\hline 
 Nd1  &  $4g$  & 0.2313 & 0.7687 & 0      & & 0.2313 & 0.7687 & 0      \\ 
 Nd2  &  $4f$  & 0.3570 & 0.3570 & 0      & & 0.3585 & 0.3585 & 0      \\ 
 Fe1  & $16k$  & 0.0373 & 0.3599 & 0.3239 & & 0.0379 & 0.3587 & 0.3237 \\ 
 Fe2  & $16k$  & 0.0675 & 0.2754 & 0.1270 & & 0.0671 & 0.2765 & 0.1269 \\ 
 Fe3  &  $8j$  & 0.0980 & 0.0980 & 0.2950 & & 0.0979 & 0.0979 & 0.2951 \\ 
 Fe4  &  $8j$  & 0.3180 & 0.3180 & 0.2542 & & 0.3174 & 0.3174 & 0.2535 \\ 
 Fe5  &  $4e$  & 0      & 0      & 0.1143 & & 0      & 0      & 0.1144 \\ 
 Fe6  &  $4c$  & 0      & 1/2    & 0      & & 0      & 1/2    & 0      \\ 
 B    &  $4f$  & 0.1236 & 0.1236 & 0      & & 0.1243 & 0.1243 & 0      \\ 
\end{tabular} 
\footnotetext[1]{Reference~\cite{li_96}}\\ 
\end{ruledtabular} 
\end{table} 
 
{\color{black} 
On the contrary, the ordered magnetic moment of each element 
varies with $U$ (Table~\ref{table:mag}). 
More correctly, the introduction of $U=5~$eV reduces $M_{\rm Nd}$ by 10\%, 
while the change in $M_{\rm Fe}$ is about 1\%. 
The magnitude of $M_{\rm Fe}$ at each site is comparable to the reported ones (see Table~\ref{tab:MFe}). 
This indicates the importance of the magnitude of $U$ for estimating $M_{\rm Nd}$ 
by DFT calculations.

\begin{table}[htp] 
\caption{\label{table:mag} 
The ordered magnetic moment of each element in ${\rm Nd_2Fe_{14}B}$ 
predicted by DFT calculations without and with $U=5~$eV.
} 
\begin{ruledtabular} 
\begin{tabular}{ccccccc} 
 &       & GGA & &   GGA+$U$ & \\ 
 & site & $M$ ($\mu_{\rm B}$) & &  $M$ ($\mu_{\rm B}$) & \\ 
\hline 
 Nd1  &  $4g$  & 2.92 &  & 2.74 &\\ 
 Nd2  &  $4f$  & 3.01 &  & 2.72 & \\ 
 Fe1  & $16k$  & 2.25 & & 2.28 & \\ 
 Fe2  & $16k$  & 2.20 & & 2.22 & \\ 
 Fe3  &  $8j$  & 2.09  & & 2.17 & \\ 
 Fe4  &  $8j$  & 2.68  & & 2.68 & \\ 
 Fe5  &  $4e$  & 2.03 & & 2.03 & \\ 
 Fe6  &  $4c$  & 2.32  & & 2.36 & \\ 
 B    &  $4f$  & -0.25 &  & -0.26 & \\ 
\end{tabular} 
\end{ruledtabular} 
\end{table} 
}

\section{\label{sec:dis}Discussion} 
\subsection{\label{sec:Nd}Nd$_2$Fe$_{14}$B} 
 
\begin{figure*}[tb] 
	\begin{centering} 
	\includegraphics[width=1.8\columnwidth]{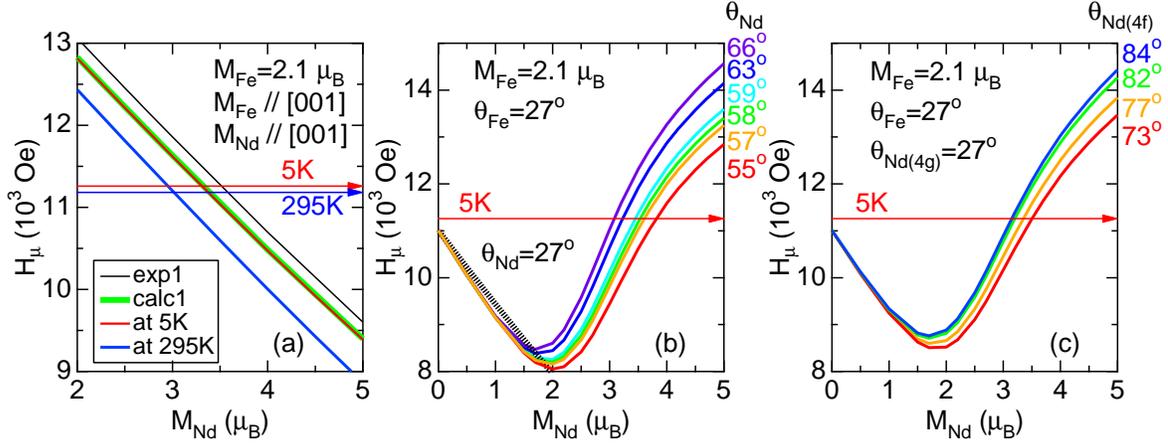} 
	\caption{ 
	The relationship between the calculated $H_{\mu}$ 
        and $M_{\rm Nd}$ in Nd$_2$Fe$_{14}$B 
	using the model that 
	(a) ${\bm M}_{\rm Fe}\parallel [001]$ 
        and ${\bm M}_{\rm Nd}\parallel [001]$;  
	(b) ${\bm M}_{\rm Fe}$ and ${\bm M}_{\rm Nd}$ 
        are both canted from the [001] direction to the [110] direction 
	with a canting angle ($\theta$) of 27$^{\circ}$ for Fe 
        and 55-66$^{\circ}$ for Nd; and 
	(c) $\theta=27^{\circ}$ for Fe and Nd at the $4g$ site, 
        but $\theta = 73$-$84^{\circ}$ for Nd at the $4f$ site. 
	In (b), a collinear FM spin arrangement ---  
        {\it i.e.} $\theta_{\rm Fe}= \theta_{\rm Nd}=27^{\circ}$ --- 
        is also shown with a broad black line. 
	} 
	\label{fig:Nd} 
	\end{centering} 
\end{figure*} 
 
For non-magnetized ferromagnetic materials in zero applied field, 
the internal magnetic field at a muon site (${\bm H}_{\mu}$) 
is represented by \cite{barth_86,schenck,yaouanc,jun_12b}; 
\begin{eqnarray} 
 {\bm H}_{\rm FM} &=& {\bm H}_{\mu} \cr 
&=& {\bm H}_{\rm dip} + {\bm H}_{\rm L} + {\bm H}_{\rm hf} \; , 
\label{eq:Hint} 
\end{eqnarray} 
This field is connected to the muon-spin precession frequencies 
through the muon gyromagnetic ratio 
[$f=H \gamma_{\mu}/(2\pi)=0.013553~$(MHz/Oe)$\times H~$(Oe)] 
leading to 
\begin{eqnarray} 
 f_{\rm FM} &=&  f_{\mu} \cr 
 &=& f_{\rm dip} + f_{\rm L} + f_{\rm hf} 
\label{eq:Fint} 
\end{eqnarray} 
where ${\bm H}_{\rm dip}$ is the dipolar field, 
${\bm H}_{\rm L}$ is the Lorentz field,  
${\bm H}_{\rm hf}$ is the hyperfine field, and 
$f_{\mu}$, $f_{\rm L}$, and $f_{\rm hf}$ are 
the corresponding muon spin precession frequencies. 
Furthermore, ${\bm H}_{\rm L}$ and ${\bm H}_{\rm hf}$ 
are connected to the saturated magnetization (${\bm M}_{\rm s}$) 
and the local spin density at the muon sites ($\rho_{\rm spin}$) 
as follows: 
\begin{eqnarray} 
{\bm H}_{\rm dip} &=& 
 -\frac{1}{4\pi\mu_0} {\bm \nabla}\left(\frac{{\bm m}\cdot{\bm r}}{r^3}\right), 
 \cr 
{\bm H}_{\rm L} &=& 
 \frac{4\pi}{3}\times {\bm M}_{\rm s}, 
 \cr 
 {\bm H}_{\rm hf} &=& 
 \frac{8\pi}{3}\times\rho_{\rm spin}({\bm r}_{\mu}). 
\label{eq:Hint2} 
\end{eqnarray} 
In order to estimate ${\bm H}_{\rm dip}$ $(f_{\rm dip})$, 
we use the results of neutron diffraction\cite{herbst_84} and 
M$\ddot{\rm o}$ssbauer\cite{rosenberg_85} measurements 
for the magnitude and direction of the Fe moments. 
Assuming that the magnitude of 
the ordered $M_{\rm Fe}$ is $2.1~\mu_{\rm B}$ \cite{herbst_91}, 
${\bm H}_{\rm dip}$ at the muon site is easily calculated 
as a function of the Nd moment using 
crystal structural data with {\sf dipelec} \cite{kojima}. 

We start by considering a collinear FM structure along the $c$-axis, 
that is, ${\bm M}_{\rm Fe}\parallel [001]$ and 
${\bm M}_{\rm Nd}\parallel [001]$. 
Since $4\pi M_{\rm s}=18.5~$kOe at 5~K 
(see Table~\ref{tab:R2Nd14B}) \cite{herbst_91,niedermayer_91}, 
${\bm H}_{\rm L}=(0, 0, 6.2$~kOe) from Eq.~(\ref{eq:Hint2}). 
Moreover, ${\bm H}_{\rm hf}=(0, 0, 0)$ 
because of the absence of any local spin density at the muon site. 
Consequently, we obtain the relationship between 
$|{\bf H}_{\mu}| = H_{\mu}^{\rm calc}$  
and the magnitude of the Nd moment ($M_{\rm Nd}$), 
as seen in Fig.~\ref{fig:Nd}(a). 
Here, the measured value of $f_{\mu}$ 
($f_{\mu}^{\rm exp}$) is 152.6(2)~MHz at 2.2~K, 
which is very close to the reported value (156~MHz) at 5~K. 
Thus, in order to explain $H_{\mu}^{\rm exp}$, 
$M_{\rm Nd}$ is uniquely determined as {\color{black}3.31}~$\mu_{\rm B}$.  
This is almost equivalent to $M_{\rm Nd}$ estimated from 
magnetization measurements, {\it i.e.} 
$M_{\rm Nd}=3.2~\mu_{\rm B}$ \cite{herbst_91}, 
confirming the reliability of the predicted muon site 
from DFT calculations. 
From the data at room temperature, {\it i.e.} 
$4\pi M_{\rm s}=16.0~$kOe at 295~K and 
$H_{\mu}^{\rm exp}=151(2)~$MHz at 300~K, 
we also obtain that $M_{\rm Nd}=3.01~\mu_{\rm B}$. 
 
\begin{figure}[tb] 
	\begin{centering} 
	\includegraphics[width=0.8\columnwidth]{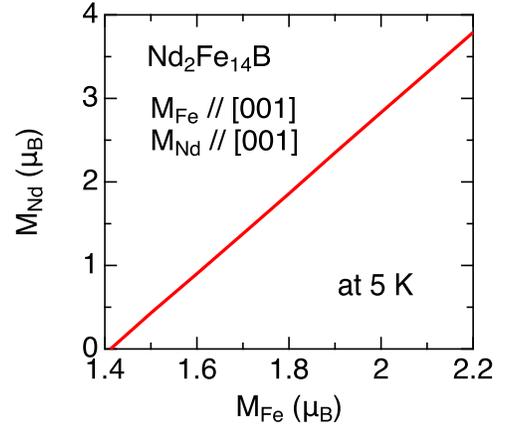} 
	\caption{ 
	The relationship between $M_{\rm Nd}$ 
        and $M_{\rm Fe}$ in Nd$_2$Fe$_{14}$B 
	using the model that 
	${\bm M}_{\rm Fe}\parallel [001]$ 
        and ${\bm M}_{\rm Nd}\parallel [001]$.  
		} 
	\label{fig:NdvsFe} 
	\end{centering} 
\end{figure} 

Although we assumed that $M_{\rm Fe}=2.1~\mu_{\rm B}$, 
{\color{black}
$M_{\rm Nd}$ estimated with the above procedure is found to increase linearly with $M_{\rm Fe}$ 
(see Fig.~\ref{fig:NdvsFe}). 
On the contrary, 
Fig.~\ref{fig:NdvsFe} provides an acceptable range for $M_{\rm Fe}$ as $2.0\le M_{\rm Fe}\le2.15~\mu_{\rm B}$, 
when $M_{\rm Nd}$ ranges between 3.0 and $3.5~\mu_{\rm B}$. 
Furthermore, we assumed that $M_{\rm Fe}$ is identical}
for all the Fe sites. 
{\color{black}However,}
experimental studies and DFT calculations reported that 
$M_{\rm Fe}$ at each site deviates slightly from $2.1~\mu_{\rm B}$. 
In order to know the effect of such deviations 
on the estimation of $M_{\rm Nd}$, 
the relationship between $H_{\mu}$ and $M_{\rm Nd}$ is also shown 
for the two cases in Fig.~\ref{fig:Nd}(a) 
and six cases in Table~\ref{tab:MFe}. 
This indicates that the four estimations for $M_{\rm Fe}$, 
{\it i.e.} exp2, exp3, calc2, and calc3, 
provide unusually large $M_{\rm Nd}$ 
under the collinear FM structure along the $c$-axis. 
 
By contrast, at low temperatures the spin orientation 
is reported to change from the [001] to the [110] direction 
below $T_{\rm SRT}=135~$K \cite{givord_84,abache_85,sagawa_85,hirosawa_86}. 
The corresponding anomaly is clearly seen 
in the $f_{\rm FM}(T)$ and $\lambda_{\rm FM}(T)$ curves 
[Fig.~\ref{fig:ZFana}]. 
More correctly, 
both Fe and Nd moments are thought to be 
{\sl canted\/} towards the [110] direction from the [001] direction,  
based on both first principles calculations 
and Fe $K$-edge x-ray magnetic circular dichroism 
(XMCD) measurements \cite{chaboy_98}. 
The canting angle ($\theta$) was estimated 
to be 27$^{\circ}$ for Fe ($\theta_{\rm Fe}=27^{\circ}$) and 
58$^{\circ}$ for Nd ($\theta_{\rm Nd}=58^{\circ}$) at 4.2~K. 
Figure~\ref{fig:Nd}(b) shows the relationship between 
$H_{\mu}$ and $M_{\rm Nd}$ for several $\theta_{\rm Nd}$ values. 
The $\mu^+$SR result clearly excludes a collinear structure, 
in which $\theta_{\rm Fe}=\theta_{\rm Nd}=27^{\circ}$, 
as an FM ground state. 
On the other hand, non-collinear structures 
provide a more plausible $M_{\rm Nd}$, 
particularly when $\theta_{\rm Nd}\sim60^{\circ}$. 
If we assume that $M_{\rm Nd}=3.2~\mu_{\rm B}$, 
$\theta_{\rm Nd}$ should be 63$^{\circ}$, 
which is very close to the value reported by XMCD (58$^{\circ}$). 

{\color{black}
Dipole field calculations provide that 
the magnetic anisotropy at the muon site ($\Theta$) is 16~deg at temperatures below $T_{\rm SRT}$, 
while $\Theta=0~$deg at temperatures above $T_{\rm SRT}$. 
Making comparison with the experimental result [$\Theta(300~{\rm K})=7(4)$deg and $\Theta(2~{\rm K})=6(4)$deg], 
the experimental accuracy of $\Theta$ was likely to be overestimated.  
This is probably due to the fact that ${\bm S}_\mu(0)$ for NSR mode is deviated from the $z$ direction 
by about 10~deg to eliminate the other particles in the muon beam. 
Nevertheless, we should note that the above estimation for $M_{\rm Nd}$ 
is based only on the magnitude of $f_{\mu}$, 
and as a result, the estimated value is not affected by the alignment of the sample.
} 
 
\begin{figure*}[htb] 
	\begin{centering} 
	\includegraphics[width=1.8\columnwidth]{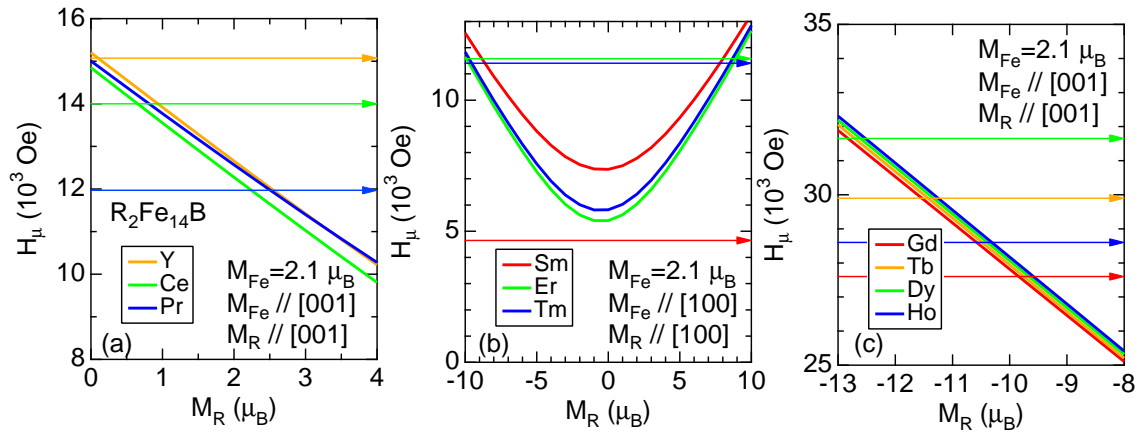} 
	\caption{ 
	The relationship between the calculated 
        $H_{\mu}$ and $M_{\rm Nd}$ in $R_2$Fe$_{14}$B 
	using the model that 
	(a) ${\bm M}_{\rm Fe}\parallel [001]$ 
        and ${\bm M}_{R}\parallel [001]$, 
	(b) ${\bm M}_{\rm Fe}\parallel [100]$ 
        and ${\bm M}_{R}\parallel [100]$, and 
	(c) ${\bm M}_{\rm Fe}\parallel [001]$ 
        and ${\bm M}_{R}\parallel [001]$.  
	In (a)-(c), the magnitude of $M_{\rm Fe}$ 
        is assumed to be $2.1~\mu_{\rm B}$. 
	}  
	\label{fig:R} 
	\end{centering} 
\end{figure*} 
 
Another XMCD study at low temperatures \cite{bartolome_00} 
proposed the possibility of a non-collinear spin arrangement 
among the Nd moments. 
That is, $\theta_{\rm Nd}\sim25^{\circ}$ for the Nd ions at the $4g$ site, 
but $\theta_{\rm Nd}\sim80^{\circ}$ for the Nd ions at the $4f$ site. 
Figure~\ref{fig:Nd}(c) shows the dependence of $H_{\mu}$ on $M_{\rm Nd}$ 
as $\theta_{\rm Nd(4f)}$ changes from 73 to 84$^{\circ}$. 
The calculations also predict that 
$\theta_{\rm Nd}=82^{\circ}$ for $M_{\rm Nd}=3.2~\mu_{\rm B}$, 
which looks consistent with the proposed arrangement. 
However, we should note that there are 
eight crystallographically equivalent muon sites ($8i$) 
in the Nd$_2$Fe$_{14}$B lattice. 
Moreover, such a non-collinear spin arrangement among the Nd moments 
produces two different $H_{\mu}$s at each $8i$ site --- 
namely, $H_{\mu}=11270~$Oe for four of the sites 
and 11655~Oe for the other four sites. 
Although the difference of the two $H_{\mu}$s (about 4\%) 
is too small to observe two distinct muon precession frequencies 
in the ZF-$\mu^+$SR spectrum, 
such a split naturally increases the field distribution width, 
resulting in an increased relaxation rate $\lambda_{\rm FM}$. 
In reality, $\lambda_{\rm FM}$ and 
$\lambda_{\rm FM}/f_{\rm FM}$ at 2~K 
are {\sl smaller\/} than those at room temperature 
[Fig.~\ref{fig:ZFana}(b)]. 
This clearly excludes the model of 
a non-collinear spin arrangement among the Nd moments 
from the FM ground state for Nd$_2$Fe$_{14}$B. 
Since the $\lambda_{\rm FM}(T)$ curve 
exhibits a broad maximum at around $T_{\rm SRT}$ 
[see Fig.~\ref{fig:ZFana}(b)], 
such a non-collinear spin arrangement among the Nd moments 
could appear in a limited temperature range 
particularly below the vicinity of $T_{\rm SRT}$. 
 {\color{black}
Even for this case, the predicted $\Theta$ is the same to that for the collinear spin arrangement among the Nd moments, 
i.e. 16~deg. 
Therefore, 
$\Theta$ provides no crucial information on the spin arrangement in Nd$_2$Fe$_{14}$B 
within the present accuracy.  
} 

\begin{table*}[hptb] 
\caption{\label{tab:MFe} 
The Fe moment at each site in Nd$_2$Fe$_{14}$B and 
the Nd moment ($M_{\rm Nd}$) estimated from the $\mu^+$SR data. 
} 
\begin{ruledtabular} 
\begin{tabular}{ccccccccc} 
case & $16k1$ & $16k2$ & $8j1$ & $8j2$ & $4e$ & $4c$ & average & $M_{\rm Nd}$   \\ 
\hline 
average & 2.1 & 2.1 & 2.1 & 2.1 & 2.1 & 2.1 & 2.1 & {\color{black}3.31} \\ 
exp1\cite{fruchart_87} & 2.08 & 2.16 & 2.06 & 2.43 & 2.28 & 1.97 & 2.16 & 3.52 \\ 
exp2\cite{onodera_87} & 2.24 & 2.30 & 2.21 & 2.55 & 2.00 & 2.17 & 2.28 & 4.27 \\ 
exp3\cite{vannoort_86} & 2.27 & 2.41 & 2.19 & 2.70 & 2.20 & 2.10 & 2.38 & 4.61 \\ 
calc1\cite{jaswal_90} & 2.15 & 2.18 & 2.12 & 2.74 & 2.13 & 1.59 & 2.20 & 3.33 \\ 
calc2\cite{hummler_96} & 2.22 & 2.28 & 2.67 & 2.16 & 1.96 & 2.43 & 2.29 & 3.80 \\ 
calc3\cite{moriya_09} & 2.28 & 2.37 & 2.32 & 2.74 & 2.19 & 2.46 & 2.38 & 4.27 
\end{tabular} 
\end{ruledtabular} 
\end{table*} 
 
\subsection{\label{sec:R}$R_2$Fe$_{14}$B}

Although we have measured $\mu^+$SR spectra only for Nd$_2$Fe$_{14}$B, 
both ${\bm H}_{\mu}$ and ${\bm M}_{\rm s}$ were reported 
for the other $R_2$Fe$_{14}$B compounds 
with $R=$ Y, Ce, Pr, Sm, Gd, Tb, Dy, Ho, Er, and Tm 
(see Table~\ref{tab:R2Nd14B}) \cite{yaouanc_87,niedermayer_91}. 
Since $4f$ electrons are well localized at the $R$ site, 
it is reasonable to assume the same muon site in $R_2$Fe$_{14}$B 
as in Nd$_2$Fe$_{14}$B. 
Concerning the spin arrangement in the FM phase, 
the easy direction of magnetization 
at base temperature \cite{herbst_91} 
revealed that both ${\bm M}_{\rm Fe}$ and ${\bm M}_{R}$ 
are parallel to the [001] direction in $R_2$Fe$_{14}$B 
with $R=$ Y, Ce, Pr, Nd, Gd, Tb, Dy, and Ho, 
but they are parallel to the [100] direction 
in $R_2$Fe$_{14}$B with $R=$ Sm, Er, and Tm.  
We also assume that $M_{\rm Fe}=2.1~\mu_{\rm B}$ 
in $R_2$Fe$_{14}$B regardless of $R$. 
 
Using the structural data of each compound, 
Fig.~\ref{fig:R} shows the relationship between $H_{\mu}$ and $M_{R}$.  
For Y$_2$Fe$_{14}$B, $M_{\rm Y}$ is estimated to be 
almost zero (0.11~$\mu_{\rm B}$), as expected for Y$^{3+}$. 
In fact, the recent photoelectron spectroscopic analysis result 
on Nd$_2$Fe$_{14}$B \cite{wang_14,min_93} 
revealed that the valence state of Nd ions is very close to 3+, 
while there is, to our knowledge, no XPS work on Y$_2$Fe$_{14}$B. 
As the atomic number increases, 
$H_{\mu}^{\rm exp}$ decreases systematically. 
From the intersection between 
$H_{\mu}^{\rm exp}$ and $H_{\mu}^{\rm calc}$, 
$M_{\rm Ce}$ and $M_{\rm Pr}$ are estimated to be 
0.66 and 2.79~$\mu_{\rm B}$, respectively (Table~\ref{tab:R2Nd14B}).   
 
For Sm$_2$Fe$_{14}$B, Er$_2$Fe$_{14}$B, and Tm$_2$Fe$_{14}$B, 
since ${\bm M}_{\rm Fe}\parallel [100]$ 
and ${\bm M}_{R}\parallel [100]$, 
the $H_{\mu}^{\rm calc}(M_{R})$ curve exhibits a parabolic shape 
with a minimum at $M_{R}=0$ [Fig.~\ref{fig:R}(b)]. 
For Sm$_2$Fe$_{14}$B, $H_{\mu}^{\rm exp}<H_{\mu}^{\rm calc}$ 
in the whole possible range of $M_{\rm Sm}$, 
leading tentatively to $M_{\rm Sm}=0$. 
This implies that the FM spin structure is 
slightly different from the proposed one \cite{hiroyoshi_85}.  
For Er$_2$Fe$_{14}$B, and Tm$_2$Fe$_{14}$B, 
there are two intersections between 
the $H_{\mu}^{\rm exp}(M_R)$ and $H_{\mu}^{\rm calc}(M_R)$ curves. 
This means that two values are available 
for $M_{\rm Er}$ and $M_{\rm Tm}$. 
However, neutron diffraction measurements proposed that 
${\bm M}_{R}$ is antiparallel to ${\bm M}_{\rm Fe}$ 
\cite{yelon_86,davis_85,yamada_85}. 
Therefore, a negative value is selected for $M_{\rm Er}$ and $M_{\rm Tm}$, 
that is, -9.94 and -9.54~$\mu_{\rm B}$, respectively.  
 
For Gd$_2$Fe$_{14}$B \cite{herbst_91}, Tb$_2$Fe$_{14}$B \cite{herbst_93}, 
Dy$_2$Fe$_{14}$B \cite{herbst_85}, and Ho$_2$Fe$_{14}$B \cite{wolfers_90}, 
${\bm M}_{\rm Fe}\parallel [001]$, 
${\bm M}_{R}\parallel [001]$, and 
${\bm M}_{R}$ is antiparallel to ${\bm M}_{\rm Fe}$. 
Indeed, $H_{\mu}^{\rm exp}$ is reproduced only when 
$M_{R}<-9~\mu_{\rm B}$ [Fig.~\ref{fig:R}(c)]. 
As a result, we obtain that 
$M_{\rm Gd}=-9.48~\mu_{\rm B}$, 
$M_{\rm Tb}=-11.4~\mu_{\rm B}$, 
$M_{\rm Dy}=-12.6~\mu_{\rm B}$, and  
$M_{\rm Ho}=-10.3~\mu_{\rm B}$.

\begin{table*}[htp] 
\caption{\label{tab:R2Nd14B} 
The internal magnetic field detected with $\mu^+$SR \cite{niedermayer_91}, 
the saturated magnetization \cite{herbst_91}, 
the magnetic moment of $R$ ($M_R$) estimated with $\mu^+$SR ($M_R^{\mu\rm SR}$), and 
$M_R$ proposed with magnetization measurements at 4~K ($M_R^{\rm Mag}$) \cite{herbst_91}, 
and $gJ$, 
where $g$ is the Land$\acute{\rm e}$ $g$-factor and $J$ is the quantum number of the total angular momentum. 
} 
\begin{ruledtabular} 
\begin{tabular}{cccccccc} 
$R_2$Fe$_{14}$B & $H_{\mu}$~(MHz) & $H_{\mu}$~(kOe) & 3$H_{\rm L}=4\pi M_{\rm s}$~(kOe) & 
$M_{R}^{\mu\rm SR}$~($\mu_{\rm B}$) & 
$M_{R}^{\rm Mag}$~($\mu_{\rm B}$)  & $gJ$ \\ 
\hline 
$R=~$Y & 204.5 & 15.07 & 15.9 & 0.11 &  0  & ---\\ 
La & --- & --- & 14.8 & --- &  ---  & 0\\ 
Ce &189.6 & 14.0 & 14.7 & 0.66 & --- &  2.14  \\ 
Pr & 162.5 & 11.97 & 18.4 & 2.79 &  3.1 &  3.20 \\ 
Nd & 152.6 & 11.26 & 18.5 & {\color{black}3.31} & 3.2 &   3.27 \\ 
Pm & --- & --- & --- & --- &  ---  & 2.40\\ 
Sm & 63.0 & 4.65 & 16.7 &  $\sim0$ & 1.0 &  0.72 \\ 
Eu & --- & --- & --- & --- &  --- & 0 \\ 
Gd & 374.0 & 27.60 & 9.2 & -9.48 & -6.8 &  7.0  \\ 
Tb & 405.2 & 29.90 & 6.6 & -11.4 & -9.1 &  9.0  \\ 
Dy & 429.0 & 31.65 & 5.7 & -12.6 & -10.1 &  10.0  \\ 
Ho & 388.0 & 28.60 & 5.7 & -10.3 & -10.1 &  10.0 \\ 
Er & 157.2 & 11.58 & 6.6 & -9.94 & -9.3 &  9.0  \\ 
Tm & 154.6 & 11.41 & 9.2 & -9.57 & -6.7 & 7.0   \\ 
Yb & --- & --- & $\sim12$ &--- & -4.2 & 4.0  \\ 
Lu & --- & --- & 14.7 & --- & --- &   0 \\ 
\end{tabular} 
\end{ruledtabular} 
\end{table*} 
 
\begin{figure}[h] 
	\begin{centering} 
	\includegraphics[width=0.8\columnwidth]{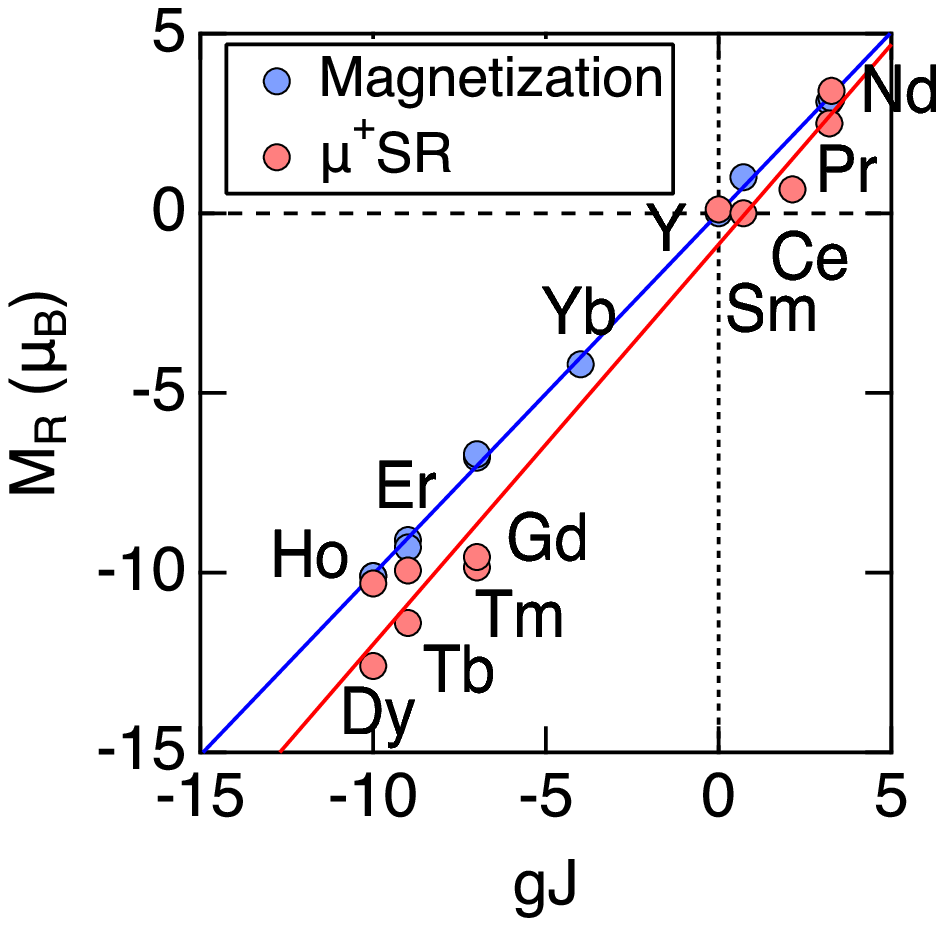} 
	\caption{ 
	The relationship between the magnetic moment of the rare earth 
        element ($M_{R}$) 
	and expected magnetic moments ($gJ$). 
	For heavy rare earth elements, negative value of $gJ$ is used, 
	because $\bm{M}_{R}$ is antiparallel to $\bm{M}_{\rm Fe}$.
	}  
	\label{fig:total} 
	\end{centering} 
\end{figure} 
 
Finally, Fig.~\ref{fig:total} shows the relationship between 
{\color{black}$M_{R}$ and the expected magnetic moment ($gJ$) 
derived from Land$\acute{\rm e}$ $g$ factor and the quantum number of the total angular momentum ($J$) 
for free $R^{3+}$ ions. 
$M_{R}$ estimated with the magnetization measurements ($M_{R}^{\rm Mag}$)  
is almost equivalent to $gJ$ \cite{herbst_91}, 
suggesting the presence of stronger exchange field to the $4f$ moments than the crystal field \cite{herbst_91}.  
On the other hand, 
the slope of the $M_{R}^{\mu\rm SR}(gJ)$ curve estimated with $\mu^+$SR 
is steeper than that for the $M_{R}^{\rm Mag}(gJ)$ curve, 
mainly because $\mid M_{R}^{\mu\rm SR}\mid>\mid M_{R}^{\rm Mag}\mid$ for the heavy rare earth elements. 
Although the reason for this discrepancy is not clear at present,  
we should note that $\mu^+$SR is very sensitive to local magnetic environments.
Recently, not only for Nd$_2$Fe$_{14}$B but also for Ho$_2$Fe$_{14}$B, 
a non-collinear spin structure for the Ho moment is proposed 
with neutron using a single crystal sample \cite{wolfers_90}. 
This implies the possibility that such non-collinear structure appears in the other $R_2$Fe$_{14}$B at low temperatures, 
which would affect the magnitude of $M_{R}^{\mu\rm SR}$. 
It would be thus an interesting subject to reconfirm the magnetic structure in $R_2$Fe$_{14}$B 
at low temperatures using a high quality sample. 
Finally,}
this work clearly demonstrates the unique power of 
a combination of $\mu^+$SR and DFT calculations 
for determining the magnetic moments of rare earth elements 
through the observation of local $H_{\rm int}$. 
 
 
\section{\label{sec:summary}Summary} 
 
We have studied the internal magnetic field in a sintered Nd$_2$Fe$_{14}$B permanent magnet sample 
with a positive muon spin rotation and relaxation ($\mu^+$SR) technique, 
{\color{black}which provides microscopic magnetic information at the muon site.} 
Combining the $\mu^+$SR data with the result of DFT calculations for predicting the muon site in the lattice, 
the magnitude of the ordered Nd moment was clearly estimated both for 
a collinear ferromagnetic structure at room temperature and 
a canted ferromagnetic structure at 2~K. 
Furthermore, a similar estimation for the ordered moment of the rare earth elements in $R_2$Fe$_{14}$B 
provided reasonable values consistent with those reported by magnetization and M\"{o}ssbauer measurements. 
{\color{black}$\mu^+$SR has been widely used for investigating a magnetic nature in  
antiferromagnetic, spin-glass, and/or paramagnetic materials, 
in which both the Lorentz field and hyperfine field are usually zero and,  
as a result, the dipole field is predominant.  
On the contrary, the present work demonstrates that a combination of $\mu^+$SR and DFT calculations further 
expands the research field into ferromagnetic materials. 
} 
 
\section{acknowledgments} 
We thank the staff of TRIUMF (especially the CMMS) 
for help with the $\mu^+$SR experiments. 
This work was supported by Japan Society for the Promotion Science (JSPS) KAKENHI Grant No. JP18H01863. 
 
 

\end{document}